\documentclass[10pt,conference]{IEEEtran}
\usepackage[pdftex]{graphicx}
\usepackage[usenames,dvipsnames,table,xcdraw]{xcolor}
\usepackage{array}
\usepackage{multirow}
\usepackage[normalem]{ulem}
\useunder{\uline}{\ul}{}
\usepackage{booktabs}
\usepackage{supertabular,booktabs}
\usepackage{longtable}
\usepackage{lscape}
\usepackage{cite}
\usepackage{url}
\usepackage{hyperref}
\usepackage{dirtytalk}
\usepackage{graphicx}
\usepackage{comment}
\usepackage{makecell}
\usepackage{svg}
\usepackage{amsmath}
\usepackage[caption=false]{subfig}
\graphicspath{{Images/}}
\usepackage[normalem]{ulem}
\useunder{\uline}{\ul}{}

\usepackage{tabularx}
\usepackage{balance}
\usepackage{nameref}

\begin{document}

\title{The Perspective of Software Professionals on Algorithmic Racism}

\author{
\IEEEauthorblockN{Ronnie de Souza Santos}
\IEEEauthorblockA{ Cape Breton University \& CESAR School \\
Sydney, NS, Canada \\
   ronnie\_desouza@cbu.ca }

\and
\IEEEauthorblockN{ Luiz Fernando de Lima }
\IEEEauthorblockA{ CESAR \\
  Recife, Brazil \\
 lffpl@cesar.org.br }
\and

\IEEEauthorblockN{ Cleyton Magalhães }
\IEEEauthorblockA{ CESAR School \\
  Recife, Brazil \\
 cvcm@cesar.school }
}


\IEEEtitleabstractindextext{%
\begin{abstract} \textit{Context}. Algorithmic racism is the term used to describe the behavior of technological solutions that constrains users based on their ethnicity. Lately, various data-driven software systems have been reported to discriminate against Black people, either for the use of biased data sets or due to the prejudice propagated by software professionals in their code. As a result, Black people are experiencing disadvantages in accessing technology-based services, such as housing, banking, and law enforcement. \textit{Goal}. This study aims to explore algorithmic racism from the perspective of software professionals. \textit{Method}. A survey questionnaire was applied to explore the understanding of software practitioners on algorithmic racism, and data analysis was conducted using descriptive statistics and coding techniques. \textit{Results}. We obtained answers from a sample of 73 software professionals discussing their understanding and perspectives on algorithmic racism in software development. Our results demonstrate that the effects of algorithmic racism are well-known among practitioners. However, there is no consensus on how the problem can be effectively addressed in software engineering. In this paper, some solutions to the problem are proposed based on the professionals' narratives. \textit{Conclusion}. Combining technical and social strategies, including training on structural racism for software professionals, is the most promising way to address the algorithmic racism problem and its effects on the software solutions delivered to our society.
\end{abstract}


}

\maketitle

\IEEEdisplaynontitleabstractindextext

\IEEEpeerreviewmaketitle

\section{Introduction}\label{sec:introduction}
Technology plays a crucial role in people’s lives, influencing several aspects of modern society, such as work, education, politics, and leisure; therefore, if software engineering does not strive to be inclusive in all its facets (i.e., education, research, and industry), software products might unintentionally constrain groups of users \cite{albusays2021diversity, adams2020diversity, fountain2022moon}. 

Currently, discussions on equity, diversity, and inclusion (EDI) \cite{mcfadden2019equity, menezes2018diversity} are gradually increasing in many fields \cite{tamtik2019policy, mcfadden2019equity}. However, in software engineering, this debate is evolving slowly \cite{myers2020discriminating}. It seems counter-intuitive, but the area responsible for creating innovative software solutions for billions of users worldwide does not reflect the diversity of the society it serves \cite{albusays2021diversity}, e.g., algorithms are racist \cite{adams2020diversity, mcfadden2019equity, owens2020those}, technical forums are sexist \cite{zolduoarrati2021value}, and the software industry is not welcoming to underrepresented groups \cite{janzen2018reflection, menezes2018diversity}.

If we analyze the discrimination disseminated by software products through the lens of racism, we can identify several examples of what is being defined as Algorithmic Racism \cite{harding2022, buolamwini2018gender}. Algorithmic racism occurs when data-drive software systems produce outcomes that reproduce racial disparities, usually against Black people \cite{milner2021data, harding2022, chung2021}, creating disadvantages in several contexts, such as housing policy, credit terms, and law enforcement. In particular, racism resulting from algorithms has intensified with the use of technology based on machine learning and artificial intelligence \cite{myers2020discriminating}.

Since software professionals have an essential role in creating technologies, in this study, we explore how these professionals understand and perceive algorithmic racism guided by the following research question:

\smallskip \smallskip
{\narrower \noindent \textit{\textbf{Research question:} What do software development professionals know about algorithmic racism?} \par}
\smallskip \smallskip

From this introduction, our study is organized as follows. In Section \ref{sec:background}, we present a literature review on algorithmic racism. In Section \ref{sec:method}, we describe how we conducted the survey. In section \ref{sec:findings}, we present our findings, which are discussed in Section \ref{sec:disussion}. Finally, in Section \ref{sec:conclusion} we present our conclusions and directions for future research. 
 
\section{Algorithmic Racism} \label{sec:background}

An algorithm is a set of instructions created by an individual to solve a problem or class of problems \cite{burkholder1987halting}. Algorithms are used along with data and statistical analyses to produce solutions; however, they do not auto-correct for human bias, e.g., biases from those who coded the algorithm or provided the data it consumes \cite{chung2021, leavy2018gender}. Therefore, unless algorithms are intentionally designed to account for discrimination by solving inequities and prejudice, they will likely reproduce and aggravate racism \cite{chung2021, koene2019governance}.

The software industry is over-represented by White heterosexual men and underrepresented by other populations, such as Black people, women, LGBTQIA+ individuals, and other equity-deserving groups \cite{myers2020discriminating, janzen2018reflection, leavy2018gender}. This representation directly affects software solutions which will carry biases from those who build them, either good or bad, conscious or unconscious \cite{harding2022, chung2021}. In other words, the underrepresentation resulting from the lack of different backgrounds in the software development allows biases to be propagated until they reach the users.

In this context, software fairness is an emerging concept in software engineering that focus on exploring strategies and techniques to develop more equitable and inclusive data-driven systems \cite{brun2018software, verma2018fairness, hort2021did}. Algorithmic racism is a concept related to software fairness as it refers to the outcomes arising from race-based bias in technology, i.e., when technology produces unequal and unfair outcomes for individuals from a specific ethnic group \cite{noble2018algorithms, kelly2021seeking}. 

Over the years, algorithms developed using machine learning approaches have been demonstrated to reproduce and intensify racial bias against Black people in several ways. For instance, for Black users, various data-driven software produce: \cite{chung2021, cheuk2021can, ghassemi2022medicine, garcia2016racist, johnson2022utilizing, sveen2022risk}: more expensive auto insurance, lower credit scores, more expensive or inaccessible mortgages, allocations for less prestigious positions, denied life-saving care, a more punitive criminal justice system, and communities over-targeted by the police.

Although the effects of algorithmic biases have affected Black people for years, the discussions on algorithmic racism began to increase only recently \cite{myers2020discriminating}, especially with the increase in the discussions on software fairness \cite{brun2018software, verma2018fairness, hort2021did} and when cases of racism resulting from software systems started to be reported by the media \cite{ormerod2022, silva2022racismo, hirota2022gender, saha2020human, russell2021machine, johnson2022utilizing, sveen2022risk}. A curious example happened when the facial recognition system used to identify suspects in the Brazilian state of Ceará added a photo of Michael B. Jordan, an African-American Hollywood actor, under the list of suspects for a mass shooting in the region \cite{ormerod2022, silva2022racismo}. 

Outside of software engineering, researchers from other areas, such as Health, Social Sciences, and Education, claim that the software industry needs to take responsibility for the effects of algorithmic racism in people's lives \cite{williams2021oh, castro2022ai+, ghassemi2022medicine, beretta2021detecting} and start to raise discussions about the risks of biased algorithms to our society \cite{hirota2022gender, saha2020human, russell2021machine, johnson2022utilizing, sveen2022risk}. Other studies concluded that discussing structural racism with software engineers is one of the leading practices to address the problem \cite{slack2020fooling, benthall2019racial, genevieve2022precision, yapo2018ethical, robinson2020teaching}. 

\section{Method} \label{sec:method}
In this study, we conducted a cross-sectional survey \cite{easterbrook2008selecting} with software professionals to explore their understanding of algorithmic racism. Our focus are algorithms created to data-driven and interactive software systems, e.g., algorithms implementing a mathematical functions are outside of the scope of this study. We followed the guidelines to build questionnaires and conduct surveys in software engineering \cite{linaker2015guidelines, pfleeger2001principles, ralph2020empirical} as described in the sections below \ref{sec:instrument}, \ref{sec:collect}, \ref{sec:analysis}, \ref{sec:ethics}. 

\subsection{Instruments} \label{sec:instrument}
We started the questionnaire by asking practitioners whether algorithms and software programs could discriminate against people. Following this, we asked them what they knew about it and if they could provide any example or experience with algorithmic racism. Next, we asked them how algorithmic racism could be avoided in software development (e.g., if they believed algorithms discriminate, how could they address the problem). Following the recommendations of \cite{baltes2022sampling}, we finished the questionnaire with demographic questions by asking participants to inform their personal and professional backgrounds.

Before collecting answers, we conducted a pilot survey with five software engineers in the industry. The pilot results were used to improve the wording, the structure of the sentences, the sequence of questions and to define how long answering the whole questionnaire would take. Table \ref{tab:questionnaire} presents the final instrument, which is organized into twelve questions divided into six sections that could be answered in up to five minutes.


\begin{table}[!htp]\centering
\caption{Survey Questionnaire}
\label{tab:questionnaire}
\scriptsize
\begin{tabularx}{\linewidth}{p{2.5cm} p{5.5cm} X}\midrule

\textbf{Section} &\textbf{Question} \\\midrule
\textbf{Consent and Participation} & 
1. This survey is COMPLETELY ANONYMOUS; no information provided can be linked back to you. Please, answer the questions with your perception and knowledge of the subject. There are no correct or incorrect answers in this survey. Answering the questionnaire will take up to 5 minutes. Do you agree to participate? \newline ( ) Yes \\ \\ 

\textbf{Algorithmic Racism Perception} & 
2. “An algorithm can produce racism.” How do you stand regarding this statement? 
\newline( ) Strongly Agree 
\newline( ) Partially Agree 
\newline( ) Undecided 
\newline( ) Partially Disagree 
\newline( ) Strongly Disagree 
\newline\newline
3. [OPTIONAL] Please, comment on your answer. \\\\

\textbf{Algorithmic Racism Background} &
4. Recently, researchers and users have been reporting ethnic disparities in the performance of systems, especially those that use machine learning. The reports say that such algorithms have discriminated against Black people in social media environments, recommendation systems, decision-making processes, and many others. Did you ever read or heard about anything linking algorithms with racism? 
\newline( ) Yes 
\newline( ) No 
\newline( ) I do not remember 
\newline\newline
5. [OPTIONAL] If you answered yes to the previous question, briefly describe the case of racism caused by algorithms that you are aware of. \\\\

\textbf{Preventing Algorithmic Racism (Open-Ended)} 
&6. How can we prevent algorithms from disseminating racism? Briefly describe what we could do to prevent algorithms from discriminating against people based on their ethnicity. \\\\

\textbf{Preventing Algorithmic Racism (Based on the Recommendations identified in the literature)} 
&7. What would be the PRIMARY method to avoid racism in algorithms? 
\newline( ) Diversifying databases and data sets 
\newline( ) Diversifying models and techniques 
\newline( ) Improving team diversity 
\newline( ) Improving system requirements 
\newline( ) Improving system design 
\newline( ) Improving testing and validation processes 
\newline( ) Training software professionals about racism 
\newline( ) Other: \\\\

\textbf{Demographics} &
8. What is your role in software development? \newline
9. What best describes your ethnicity? \newline
10. What best describes your gender? \newline
11. What best describes your sexual orientation? \newline
12. Do you identify as transgender? \\
\bottomrule
\end{tabularx}
\end{table}

\subsection{Data Collection} \label{sec:collect}
We followed the recommendations of \cite{baltes2022sampling} and applied three data collection techniques, namely, convenience sampling, purposive sampling and snowballing sampling. These techniques are examples of non-probability sampling, that is, sampling methods that do not employ randomness. In this sense, data collection happened as follows:

\begin{itemize}
\item \textit{Convenience sampling} relies on selecting participants based on their availability \cite{baltes2022sampling}. In this process, we invited software engineers from our professional network to participate in the study and answer the questionnaire according to their availability. 

\item \textit{Purposive sampling} relies on selecting participants from a specific site or source and inviting them to participate in the study \cite{baltes2022sampling}. In this process, we sent the questionnaire to software professionals from a large software company in South America who participated in our previous studies. The company has over 1200 professionals developing software solutions for several partners from North America, Asia, and Europe. South America is among the ideal regions to conduct a survey focused on algorithmic racism against Black people due to its history and the number of Afro-descendants living in South American countries, e.g., Brazil. 
\newline

\item \textit{Snowballing sampling} relies on requesting individuals from the population to identify other members that could participate in the study  \cite{baltes2022sampling}. Following this technique, we asked participants from the convenience sample to forward our questionnaire to colleagues and co-workers. 
\end{itemize}

\subsection{Data Analysis} \label{sec:analysis}
Both qualitative and quantitative data were collected with our questionnaire. Following this, our data analysis strategy was based on descriptive statistics \cite{george2018descriptive} to explore the quantitative data and thematic analysis \cite{cruzes2011recommended} to analyze the qualitative data.

Descriptive statistics \cite{george2018descriptive} allowed us to explore the participants' profiles (demographics) and organize the distribution and the frequency of their answers to closed-ended questions. In addition, it allowed us to create partitions of responses and group practitioners using different statistical and mathematical functions (e.g., means, proportions, totals, ratios) to integrate the data. We used Tableau to support the quantitative analysis and identify correlations among groups.

Thematic analysis \cite{cruzes2011recommended}, line-by-line coding and focused coding supported us in grouping the experiences and perceptions reported by participants into high-level themes that demonstrated their understanding of the problem. Through line-by-line coding, we explored the individual answer to the questionnaire and then created codes that were grouped and categorized based on the existing connections among them using focused coding. 
 
\subsection{Ethics} \label{sec:ethics}
No personal information about the participants was collected in this study (e.g., name or e-mail) to maintain participants’ anonymity. In addition, we included at the beginning of the questionnaire general information about the research team and a statement of anonymity and volunteer participation. Finally, we asked participants to agree (by checking a yes box) to use their data for scientific purposes. 

\section{Findings} \label{sec:findings}

Table \ref{tab:demographics} summarizes the profile of 73 professionals who participated in this survey. In general, our sample is very diverse, composed of professionals from different ethnicities, gender, and sexual orientation. In addition, professionals with different roles in software development participated in the survey, providing us with various professional experiences. 

Diversity is an essential factor in this study because it allows us to explore how individuals from underrepresented groups in software engineering, e.g., non-White individuals and other minorities, perceive algorithmic racism in comparison with individuals that belong to the majority group in the software industry.

\begin{table}[!htp]\centering
\caption{Participants Profile}
\label{tab:demographics}
\scriptsize
\begin{tabularx}{\linewidth}{p{3cm} p{3cm} p{5cm}} \midrule

Ethnicity
&White &48 individuals\\
&Mixed-race &18 individuals$^a$\\
&Black &6 individuals\\
&No answer &1 individual\\
\midrule

Gender 
&Male &44 individuals$^b$\\
&Female &25 individuals\\
&Non-binary &2 individuals$^b$\\
&No answer &2 individuals\\
\midrule

Sexual Orientation
&Straight &54 individuals\\
&Bisexual &7 individuals\\
&Gay &4 individuals\\
&No Answer &3 individuals\\
&Lesbian &2 individuals\\ 
&Asexual &2 individuals\\ 
&Queer &1 individual\\ 
\midrule

Role
&Programmer &30 individuals\\
&Designer &19 individuals\\
&Manager &10 individuals\\
&Tester/QA &9 individuals\\
&Data Scientist &3 individuals\\
&Requirements Analyst &1 individual\\
&No answer &1 individual\\
\bottomrule
\end{tabularx}
\flushleft
\footnotesize{Notes: $^a$Mixed-race includes people of more than ethnicity, including Afro-descendants and Indigenous-descendants; $^b$one man and one non-binary person are transgender}

\end{table}

\subsection{Algorithmic Racism: Software Professionals Understanding} \label{sec:under}

\begin{figure*}[t]
\centering
\centerline{\includegraphics[width=0.73\textwidth]{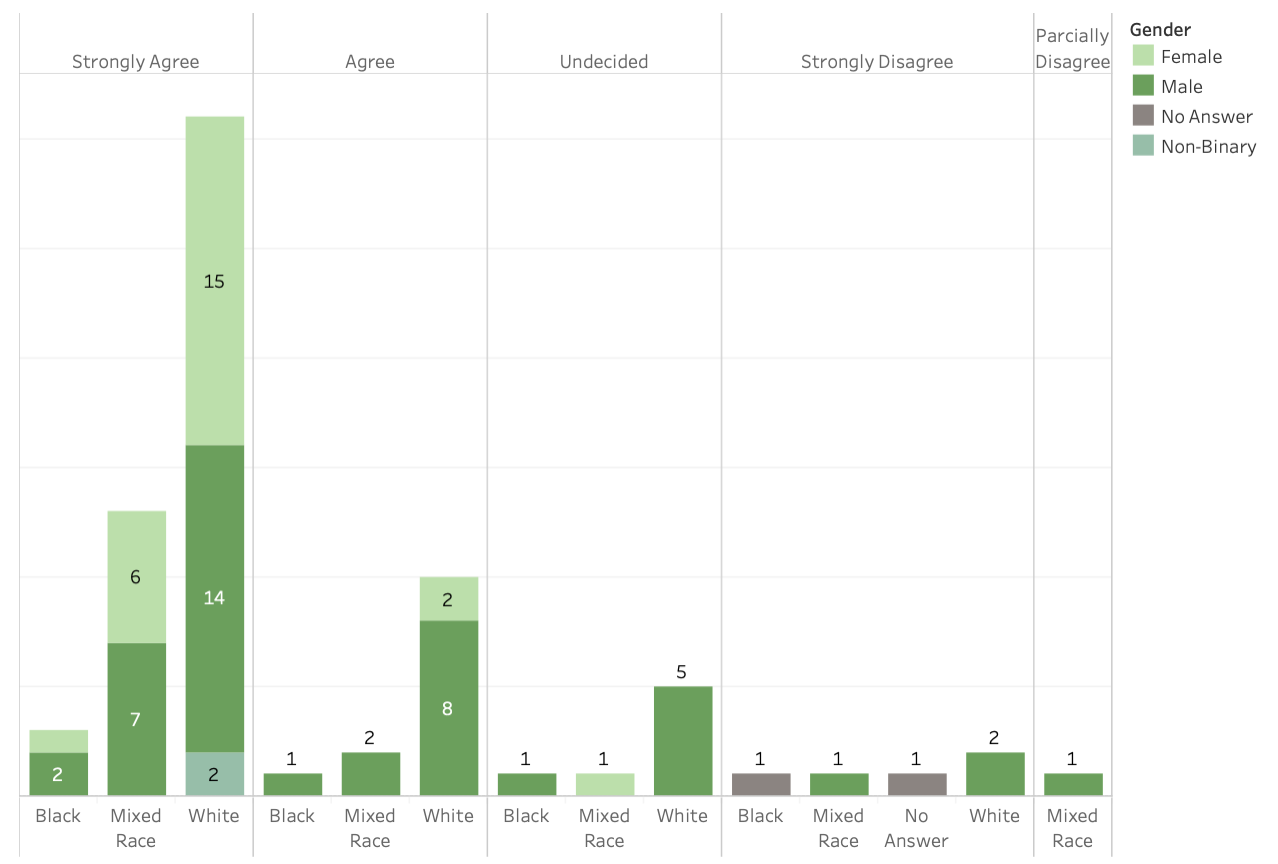}}
\caption{Algorithmic Racism Perception: Gender and Ethnicity}
\label{fig:gendereth}
\end{figure*}

\begin{figure*}[t]
\centering
\centerline{\includegraphics[width=0.73\textwidth]{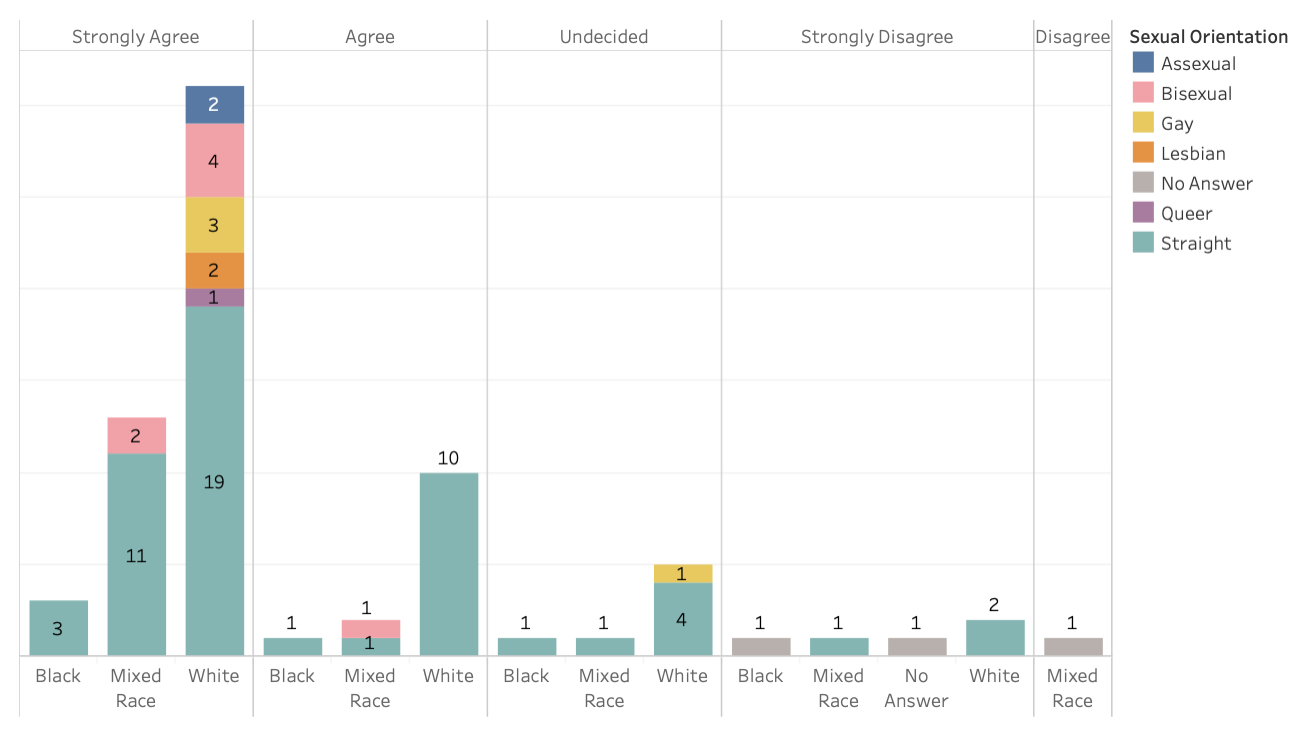}}
\caption{Algorithmic Racism Perception: Sexual Orientation}
\label{fig:diversity}
\end{figure*}

Our results show a general consensus among software professionals about algorithmic racism. Over 82\% of the sample agreed that algorithmic racism exists at some level. On the other hand, 8\% of the sample disagree that algorithms can discriminate. Less than 10\% of the sample is undecided or does not know the subject. The distribution of answers is as follows:

\begin{itemize}
\item Over 64\% of the sample strongly agree that algorithms can disseminate racism.
\newline
\item About 18\% of the sample partially agrees that algorithms can disseminate racism.
\newline
\item 9\% of the participants are undecided or have no knowledge of the subject.
\newline
\item 7\% of the sample strongly disagree that algorithms can cause damage to individuals based on their ethnicity, and 1\% only partially disagree with this.
\end{itemize}

Further analyzing the sample, we observed that the vast majority of participants from underrepresented groups agree that algorithms and systems can discriminate users based on their ethnicity, as shown in Figures \ref{fig:gendereth} and \ref{fig:diversity}. This result indicates that individuals that belong to underrepresented groups in the software industry, such as women and LGBTQIA+ people, are more empathetic to aspects related to discrimination (e.g., racism), which reinforces the importance of the debate on the role of diversity in software engineering. Therefore, our results indicate that diversity on software teams contributes to the development of algorithms that are less likely to discriminate against users based on their ethnicity. 

Regarding the role that professionals have in software development, all designers and software managers in the sample agree that algorithms can discriminate against users. On the other hand, programmers (7), software testers (3), requirements analysts (1), and data scientists (1) are among those who disagree with it or are undecided about the problem. This result indicates that technical diversity provides less awareness about algorithmic racism among the professionals. 

\subsection{Algorithmic Racism: Software Professionals Experiences} \label{sec:exper}

We asked participants to comment on the problem by describing how algorithms can discriminate against people based on their ethnicity, and 41 participants provided examples. We grouped these examples into categories and described them using the participants' descriptions. Table \ref{tab:examples} illustrates the categories with quotations extracted from the questionnaires. In summary, our results demonstrate that software professionals are aware of algorithmic racism in the following scenarios: 

\begin{itemize}
\item \textit{Image Recognition}: Twenty participants described that algorithmic racism often happens in systems that use facial recognition. According to them, these tools frequently fail to identify Black users correctly. This discrimination can cause several negative impacts and is responsible for associating Black people with criminal activities in security systems or in restricting the access of some users to various services, e.g., apps with facial validation. 
\newline

\item \textit{Biased Code}: Seven participants argued that not all programmers are prepared to deal with aspects related to racism, and therefore they can (unconsciously) propagate personal biases into their code or do not perceive biases in the data they use. In this sense, the algorithm will produce outcomes that constrain groups of users, for instance, when the data used by the algorithm is biased, but the engineer fails to address the issue. 
\newline

\item \textit{Web Searches}: Seven participants brought up examples of how web engines discriminate against non-White individuals in several ways. According to the participants, search algorithms continuously associate bad outcomes with Black people and good outcomes with White people and use this association to retrieve biased results.
\newline

\item \textit{Bad Experience}: Seven participants understand that algorithmic racism happens when systems end up prioritizing a group of people over another or when non-inclusive design produces software interactions that mistreat users and keep them from having a good experience, for instance, when the system does not offer options that match the needs of Black people. 
\end{itemize}

\begin{table}[!htp]\centering
\caption{Examples of Algorithmic Racism Perceived by Participants}
\label{tab:examples}
\scriptsize
\begin{tabularx}{\linewidth}{p{2cm} p{6cm}}

\textbf{Type} &\textbf{Quotations} \\\midrule
Image Recognition &
“Identifying Black people as criminals even when they are not or indicating them as those who are likely to commit a crime.” (P10) \newline \newline
“I had to prove to this online shopping website that I was myself because their app cannot recognize my face”. (P60) \newline \newline
“Twitter crops images based on the position of white people in the photo.” (P63) \\ \midrule

Biased Code &
“The algorithms ‘actions’ fully reflect the action of those who build them.” (P61) \newline \newline
“When the algorithm is programmed by white professionals, it is possible that this programming is done based on their own life experience and ends up excluding or differentiating patterns that are different from their own.” (P56) \newline \newline
“An algorithm is a series of calculations created by people, and there is no way for them to not apply what they think to the preferences and determinations of the algorithm.” (P31) \\ \midrule

Web Searches &
“If you google ‘handsome man’ only images of White people with blue eyes will pop up, but if you search for ‘ugly man’, it will always retrieve Black people.” (P21) \newline \newline
“Retrieving photos of people with curly hair when you search for ‘bad hair’.” (P39) \\ \midrule

User Exclusion &
“Job platforms can disadvantage resumes from individuals from underrepresented groups.” (P68) \newline \newline
“I've seen artificial intelligence generating only White people images for professional positions that require a high level of education, such as lawyers, doctors, and engineers.” (P50) \newline \newline
“When the system interaction/graphical interface is aggressive and racist to users.” (P47) \\
\bottomrule
\end{tabularx}
\end{table}

\subsection{Algorithmic Racism: Software Professionals Suggested Solutions} \label{sec:solutions}
\begin{figure*}[t]
\centering
\centerline{\includegraphics[width=0.8\textwidth]{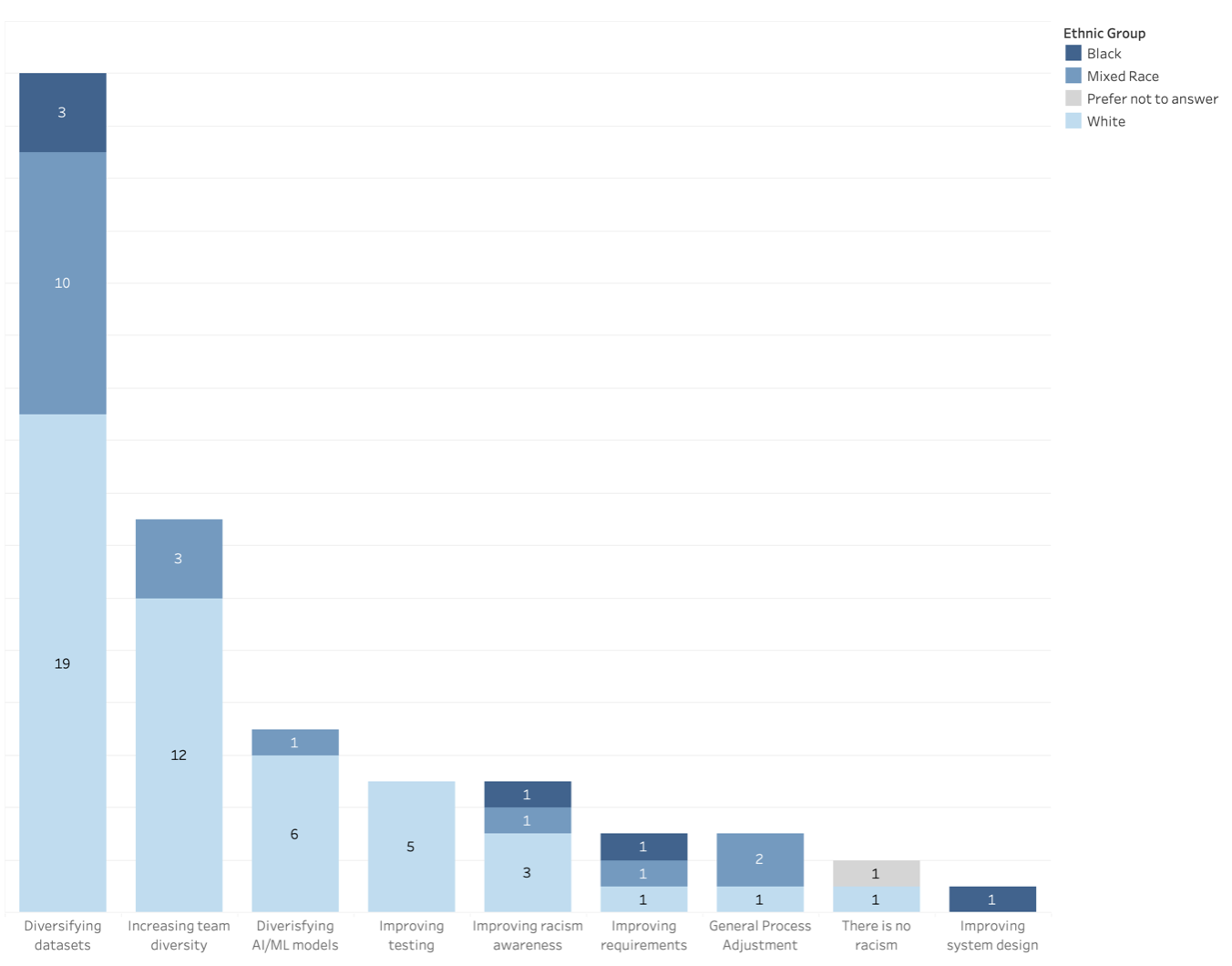}}
\caption{Algorithmic Racism Solutions: Ethnicity}
\label{fig:solueth}
\end{figure*}

\begin{figure*}[t]
\centering
\centerline{\includegraphics[width=0.8\textwidth]{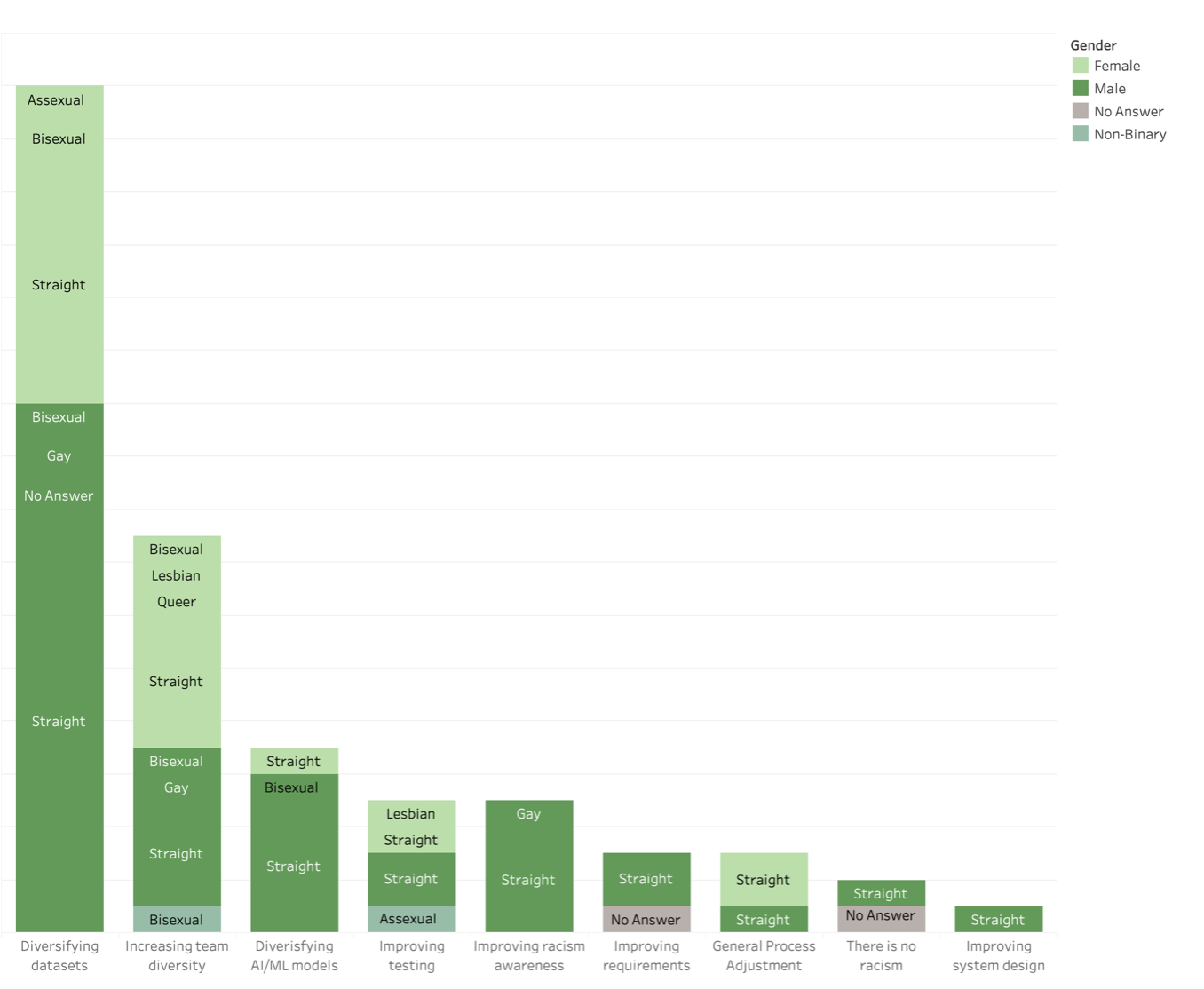}}
\caption{Algorithmic Racism Solutions: Gender and Sexual Orientation}
\label{fig:solugender}
\end{figure*}

\begin{figure*}[t]
\centering
\centerline{\includegraphics[width=0.8\textwidth]{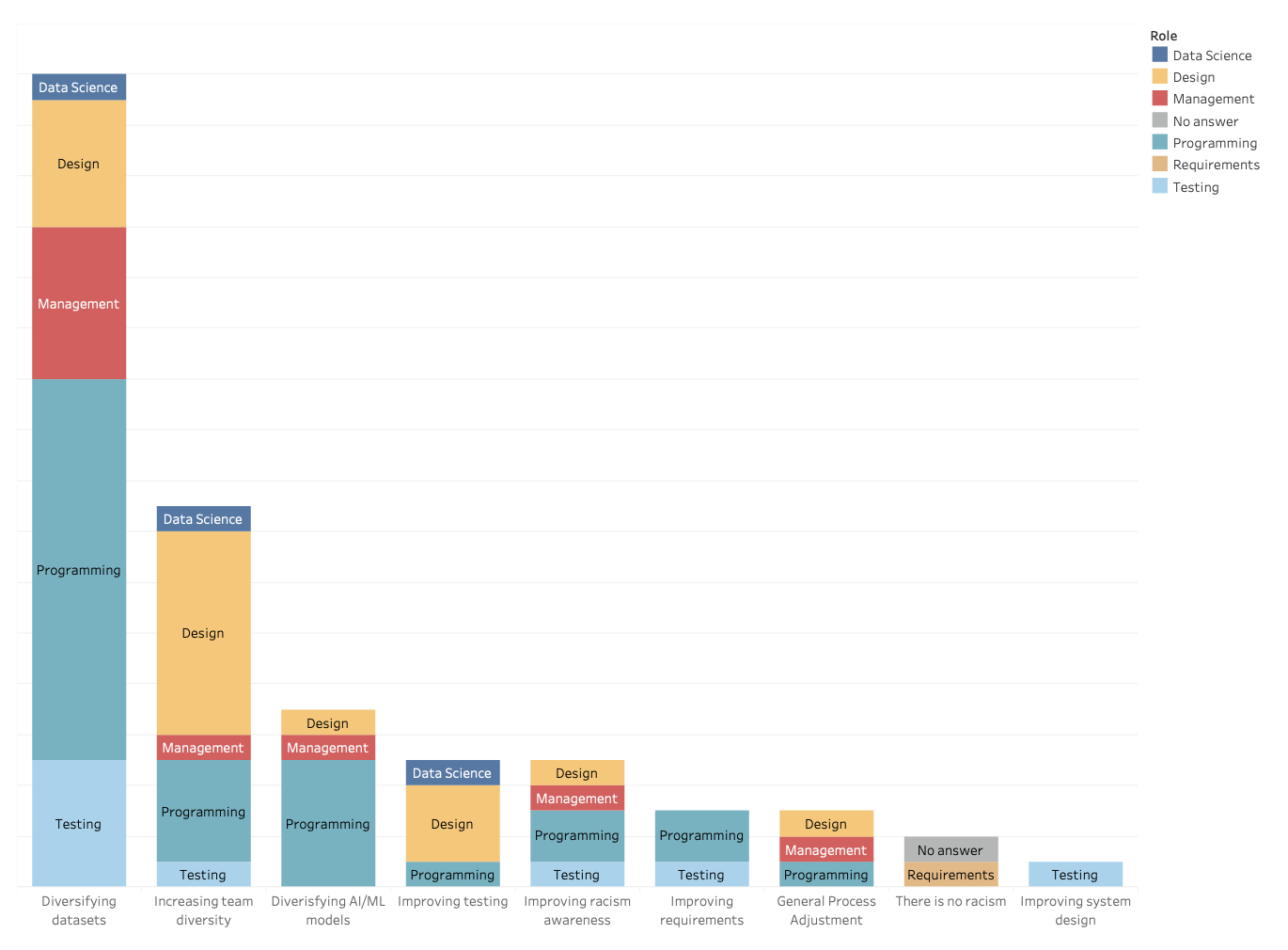}}
\caption{Algorithmic Racism Solutions: Roles}
\label{fig:soluroles}
\end{figure*}

We asked participants how to avoid or prevent algorithmic racism and built a list of recommendations proposed by them. Figure \ref{fig:solueth}, Figure \ref{fig:solugender} and Figure \ref{fig:soluroles} present the distribution of their answers considering diversity of personal background (ethnicity, gender, and sexual orientation) and diversity of technical background. Following this, Table \ref{tab:solutions} illustrates these results with quotations extracted from the questionnaires. In summary our results indicate that:

\begin{itemize}
\item Most professionals (44\%) understand that the foremost solution to this issue is diversifying data sets to generate less biased patterns.
\newline
\item Over 20\% of professionals indicate that to address racism, software teams need to become more diverse.
\newline
\item More than 9\% of participants believe that the solution for algorithmic racism must focus on diversifying the techniques (e.g., machine learning models) used to build the algorithms.
\newline
\item Almost 7\% of professionals claim that software teams must be trained in diversity and inclusion as a strategy to promote racism awareness.
\newline
\item About 7\% of participants believe that improving testing processes (e.g., fairness testing) is the primary solution to the problem. 
\newline
\item More than 5\% of participants stated that the general software processes, in particular when building machine learning algorithms, need to be revised.
\newline
\item About 4\% understand that the solution is more related to improving software requirements.
\newline
\item Finally, 3\% of professionals in the sample believe that algorithms cause no racism.
\newline
\end{itemize}

\begin{table}[!htp]\centering
\caption{Strategies to Address Algorithmic Racism}
\label{tab:solutions}
\scriptsize
\begin{tabularx}{\linewidth}{p{2cm} p{6cm}}
\textbf{Strategy} &\textbf{Quotations} \\\midrule
Diversifying Datasets &
“We can use more inclusive databases to train IA/ML algorithms.” (P22) 
“How about creating a separate dataset for cases that were considered racism.” (P35) \\

Increasing Team Diversity &
“We should increase the number of black developers not only in engineering tasks but in leadership roles where the decisions are made.” (P72) 
“Include people of diverse backgrounds, and ethnicities in the development of algorithms, especially in the data analysis, training, and model validation phases.” (P29) \\
\midrule

Diversifying AI/ML models &
“Since most of the available information is already biased, we need to define new filters for the information being fed into the algorithm." (P32) \newline\newline
“We can apply metrics to appraise the models in comparison to the differences in the outcomes they generate.” (P10) \\
\midrule

Improving Testing &
“Testing with several people who are different from each other (not only with white people).” (P67) \newline\newline
“I think we should test it with people from all possible races and not just with a few common cases.” (P69) \\
\midrule

Improving Racism Awareness &
“Provide better training in Social Sciences to Technology students to support the development of a mature and more critical view of society and the various social and cultural contexts in which we live.” (P56) \newline\newline
“Continuously spreading knowledge about racism in the technology world, which, unfortunately, is still predominantly White.” (P72) \\
\midrule

Adjusting General Processes &
“We need to understand what is causing it and work around it.” (P13) \newline\newline
“Companies that have racially biased algorithms should be held accountable.” (P29) \\
\midrule

Improving Requirements &
“I believe that the algorithm itself has to be conceived since the first step having racism in mind.” (P04) \newline\newline
“We should improve requirements and the interaction with the users.” (P42) \\
\midrule

There is no racism &
“I don’t see racism in algorithms.” (P18) \newline\newline 
“Use euphemisms if you must. Anything besides this is interfering with the data (lying). Twitter got fuc**d because of stupid activists such as the author of this survey (he/she/they/x/etc.).” (P43) \\
\bottomrule
\end{tabularx}
\end{table}

One detail was highlighted during the analysis of recommended solutions. Although six participants in the sample disagreed that algorithms could discriminate and seven participants were unsure about it, in this stage of the survey (i.e., discussing solutions for the problem), only two participants answered that there is no racism associated with algorithms, as illustrated in Figure \ref{fig:solueth} and \ref{fig:solugender}. As for the participants who answered that algorithms could not discriminate, the perception of one software professional in the survey is somehow 'extreme', as the quotations show.

Finally, exploring Figures \ref{fig:solueth} and \ref{fig:solugender}, we can observe how the perspective and the experiences of a diverse team are essential to raising solutions that can be applied to address the algorithmic racism problem. These diversity is associated to several solutions, e.g., technical, managerial and educational. 

\section{Discussion} \label{sec:disussion}
We start our discussions by comparing our results with the previous literature. Following this, we discuss implications and threats to validity. 

\subsection{Enfolding the Literature}  \label{sec:enfo}
There are a few similarities between what has been discussed in the literature over the years and the perceptions of software professionals about the theme, as both researchers and practitioners reported harms caused by algorithmic racism regarding how systems deal with images (e.g., social media) and how web engines can distort results based on ethnicity \cite{koene2019governance, noble2018algorithms, cheuk2021can}. In particular, practitioners have a strong perception of the damages caused by algorithmic racism on systems that apply facial recognition. 

Nowadays, the primary concern observed in the literature is how algorithmic racism can affect decision-making tools and disadvantage Black people by targeting, restricting, and excluding them  \cite{williams2021oh, castro2022ai+, ghassemi2022medicine, beretta2021detecting, hirota2022gender, saha2020human, russell2021machine, johnson2022utilizing, sveen2022risk}. In particular, there is a growing concern about how the biased data present in algorithms used by the police is likely to target this population at the same time that they are being excluded from public policies and having their access to essential services reduced by biased systems used in healthcare, banking, and housing. Most of these issues were not perceived or cited by practitioners.

The idea that software practitioners perceive racism more frequently in image recognition processes and web searchers might be related to the context in which they are inserted. Software professionals deal with technology daily, from conception to building and using it. It is only logical that technological issues catch their attention. This can explain why racism related to decision-making processes and Black people's exclusion from services were mainly reported from studies conducted in other areas, such as Health and Social Sciences \cite{ormerod2022, silva2022racismo, hirota2022gender, saha2020human, russell2021machine, johnson2022utilizing, sveen2022risk}. This scenario reinforces the need for raising awareness of the problem in the software industry. 

However, our study demonstrates that practitioners have a broader view of algorithmic racism than what is published in the literature so far. For instance, software professionals understand that algorithms can discriminate not just by restricting and excluding users but also when the system interface (e.g., GUI) constrain Black users (e.g., with aggressive outputs or lack of options). Again, this is a technical detail that is likely to be perceived by professionals who work with software development. 

Looking at social aspects, our study revealed that among software professionals, software designers are more likely to be aware of algorithmic racism. These are not only the professionals that provided more examples of the problem, but they are also the ones who could perceive that addressing this issue involves both the technical and social aspects, e.g., not simply exploring data sets and AI/ML models but also increasing team diversity and raising awareness about structural racism to help professionals recognizing it. On the other hand, professionals who have other roles in software development (e.g., programmers) understand that the solution lies primarily in technical factors, such as how the data sets are structured.

Practitioners agreed that the best strategy to solve algorithmic racism is by diversifying data sets. We claim that this solution is limited as it addresses only part of the problem since new data inputs might continuously be impregnated with racism. Therefore, improving software team diversity and increasing racism awareness among software professionals will likely produce long-term results. In particular, we need the software industry to engage software engineers in discussions about software fairness and its effects on technology. 

Some professionals in our sample recognized the need for more understanding about aspects related to fairness and proposed that software companies raise discussions on the subject (e.g., talks and seminars). In addition, Computer Science, Software Engineering, and other technology courses play an essential role in training students on software fairness, helping them become professionals who understand their responsibility in developing unbiased algorithms that produce a fair experience for people. 

\subsection{Implications} \label{sec:implications}

This study has implications for industrial practice. Our results demonstrated that software professionals are aware of algorithmic racism and the harm it causes to individuals in our society. However, addressing this problem requires software companies to explore not just technical aspects of software development but also work towards improving awareness of structural racism among professionals and increasing the diversity of software teams. 

Our study demonstrated how important it is to inform software professionals about their responsibility in building technologies that are inclusive and unbiased. As per our results, the number of participants in this survey that disagreed that algorithms could discriminate is small. However, among those who disagreed, some have an 'extreme' perception of the subject (e.g., Participant 43 - \ref{tab:solutions}). Software companies have a social obligation to protect users from these types of professionals and their intolerance. Other areas and industries that consume software (e.g., Health and Law) have already expressed their concerns about this. 

Therefore, based on the obtained results, we recommend that software companies:

\begin{itemize}
\item Provide training to software professionals about the impact of technology on society, including structural racism and how it can affect algorithms and software.
\newline
\item Establish strategies to deal with biased data sets to identify possible sources of racism that can be disseminated into data-drive software systems.
\newline
\item Improve the software development process by incorporating methodologies and practices that focus on fairness (e.g., inclusive design, and fairness testing) to minimize risks of discrimination in software products.
\newline
\item Increase diversity in software teams by having professionals from different backgrounds working on software solutions, as various experiences increase the perception on biases and discrimination.
\newline
\item Establish equity, diversity, and inclusion committees composed of software professionals from different backgrounds (e.g., ethnicity, gender, and sexual orientation) that could increase discussions on the theme in the organization.
\newline
\item Publicly declare the company’s position against racism and discrimination to discourage such attitudes among employees and reduce the chances of individual prejudices being reflected in software products.
\end{itemize}

Finally, our study raises several opportunities for researchers to continue investigating the topic, for instance: a) replicating this survey with professionals from specific technical backgrounds that are underrepresented in our sample (e.g., requirements and testing); b) comparing the perspective of users and professionals about the theme; c) applying scientific interventions to address the problem (e.g., action research); d) investigating algorithmic racism through the lenses of policies and regulations about ethics in artificial intelligence that are currently being developed in many countries.

\subsection{Threats to Validity} \label{sec:limitations}
Cultural aspects are the main limitation of the study. The participants' perceptions are influenced by the environment where they live and work. As we started with convenience sampling in a large company from South America, cultural aspects of the region are associated with our results. However, we believe that investigating racism in a region where the population is diverse resulted in interesting findings about the problem. Of course, we cannot claim generalization of results because racism itself is an attitude that changes depending on the region where it is investigated, e.g., underrepresented groups in South America are likely to be different from underrepresented groups in East Asia, and so on. Our results are context dependent as they focus on racism involving Black people, and although they are not generalizable, we expect they can be re-analyzed and considered for discussions in other contexts.

\section{Conclusion} \label{sec:conclusion}
We conducted a cross-sectional survey to explore the understanding of software professionals on how algorithms disseminate racism against Black people. We obtained answers from a diverse sample of 73 software professionals that discussed examples of algorithmic racism, commented on their experiences about it, and suggested solutions on how to address the problem. 

Our results demonstrated that algorithms discriminate against Black people in many ways, in particular when they use machine learning techniques. Biases in data sets have produced outcomes that restrict and exclude individuals from essential services, while in other contexts, such as the criminal justice system, algorithms have unfairly targeted this population. 

Several strategies can be used to address the problem, including diversifying data sets, using different machine-learning models, and improving testing fairness. However, the most effective solution is combining these technical approaches with strategies supported by social aspects, such as promoting structural racism training and increasing diversity in software teams. 

Since the role of software engineering is to provide an equal and fair experience to users, the software industry needs to start addressing racism, not only by finding solutions for the algorithmic biases but also by fostering opportunities for professionals from underrepresented groups as a strategy to incorporate different perspectives into software products and improve software fairness. 

Similarly, researchers are expected to address the problem by investigating how biased algorithms affect society in several contexts (e.g., health, education, justice, etc.) and how they can be improved to be more inclusive. This is a research effort that cannot be limited only to the software industry, and it needs to involve software engineering education. 

For future studies, we expect to explore further the perspective of software professionals on algorithmic racism identified in this study by conducting qualitative studies, such as case studies and grounded theory. Finally, we plan to identify collaborations to replicate the survey in other contexts. 



\ifCLASSOPTIONcaptionsoff
  \newpage
\fi

\balance
\bibliographystyle{IEEEtran}
\bibliography{bib.bib}

\end{document}